\documentclass[11pt]{article} 
 
\usepackage{amssymb,cite} 
\usepackage{graphicx} 
\usepackage{latexsym} 
\usepackage{epsfig,epic,amsfonts} 
\usepackage{wrapfig} 
\def\thefootnote{\fnsymbol{footnote}} 
 
\newcommand{\eq}{\begin{equation}} 
\newcommand{\en}{\end{equation}} 
\newcommand{\be}{\begin{equation}} 
\newcommand{\ee}{\end{equation}} 
\newcommand{\eqa}{\begin{eqnarray}} 
\newcommand{\ena}{\end{eqnarray}} 
\newcommand{\ba}{\begin{eqnarray}} 
\newcommand{\ea}{\end{eqnarray}}

\newcommand{\ZZ}{\hbox{{\rm Z{\hbox to 3pt{\hss\rm Z}}}}}

\newcommand{\EQ}{\begin{equation}} 
\newcommand{\EN}{\end{equation}} 
\newcommand{\bea}{\begin{eqnarray}} 
\newcommand{\eea}{\end{eqnarray}}

\begin{document} 
\begin{titlepage} 
\vskip0.5cm 
\begin{flushright} 
DFTT 11/06\\ 
gef-th-13/06\\ 
PTA/06-17\\ 
\end{flushright} 
\vskip0.5cm 
\begin{center} 
{\Large\bf  Study of the flux tube thickness in 3d LGT's by means of 2d spin models.} 
\end{center} 
\vskip1.3cm 
 
\centerline{    
M. Caselle$^{a}$, P. Grinza$^{b}$ and N.    
Magnoli$^{c}$}    
    
 \vskip0.4cm    
 \centerline{\sl  $^a$ Dipartimento di Fisica    
 Teorica dell'Universit\`a di Torino and I.N.F.N.,}    
 \centerline{\sl via P.Giuria 1, I-10125 Torino, Italy}    
 \centerline{\sl    
e--mail: \hskip 1cm    
 caselle@to.infn.it}    
 \vskip0.1 cm    
 \centerline{\it $^{b}$ Laboratoire de Physique Th\'eorique et Astroparticules,    
Universit\'e Montpellier II,}    
 \centerline{ Place Eug\`ene Bataillon, 34095    
Montpellier Cedex 05, France}    
 \centerline{\sl    
e--mail: \hskip 1cm grinza@lpta.univ-montp2.fr}    
 \vskip0.1 cm    
 \centerline{\sl $^e$  Dipartimento di Fisica,    
 Universit\`a di Genova and    
 I.N.F.N.,}    
 \centerline{\sl via Dodecaneso 33, I-16146 Genova, Italy}    
 \centerline{\sl    
e--mail: \hskip 1cm magnoli@ge.infn.it}    
 \vskip0.4cm    
    
\begin{abstract}    
We study the flux tube thickness in the confining phase of the $(2+1)d$ $SU(2)$ Lattice Gauge Theory near  
the deconfining phase transition.  
Following the Svetitsky-Yaffe conjecture, we map the problem to the study of the $ <\epsilon\sigma\sigma>$  
correlation 
function in the two-dimensional spin model with $Z_2$ global symmetry, (i.e. the $2d$ Ising model) 
in the high-temperature phase. Using the form factor approach we obtain an explicit expression for this function 
and from it we infer the behaviour of the flux density of the original  $(2+1)d$ LGT. Remarkably enough the 
result we obtain for the flux tube thickness agrees  
(a part from an overall normalization) with the effective string prediction for the same quantity. 
\end{abstract}    
\end{titlepage}    
 
\setcounter{footnote}{0} 
\def\thefootnote{\arabic{footnote}} 
\section{Introduction} 
\label{introsect}

The distinctive feature of the interquark potential in a confining gauge theory is that the  
colour flux is confined into a thin flux tube, joining the quark-antiquark pair. As it is well known 
the 
quantum fluctuations of this flux tube (which are assumed to be described by a suitable effective string model)  
lead to a logarithmic increase of the width of the flux tube as a function of the interquark distance $R$. This 
behaviour was discussed many years ago by L\"uscher, M\"unster and Weisz in~\cite{lmw80} and is one of the 
most stringent predictions of the effective string description of confining LGT's.  
Indeed for a non-confining theory one would instead  
expect a linear increase of the width of the flux tube. 
 
A natural question is what happens of this picture at the deconfinement point. One would naively expect a sudden 
jump of the flux tube thickness from a log to a linear dependence from the interquark distance. However we shall 
show in this paper that this is a 
misleading picture. Indeed the flux tube width also depends on the finite temperature of the theory. In the 
standard finite temperature setting of LGT's in which the quarks are represented by Polyakov loops this means that 
the flux tube thickness (and its $R$ dependence)  
also depends on the lattice size in the "time-like" compactified direction.  
 
A tentative answer to this question can be obtained in the effective string framework.  
As we shall see in sect.2.4 by using a duality transformation it is possible to show that as the temperature 
increases the log behaviour smoothly moves to a linear behaviour, thus excluding a log to linear transition at 
the deconfinement point. However this result strongly relies on the effective string approximation (even worse 
on the gaussian limit of the effective string) and it would 
be nice to have some kind of independent evidence. 
 
Unfortunately  the flux tube thickness (and in particular its dependence on the interquark distance) 
is very difficult to study by Montecarlo simulations. The only existing  
numerical estimates are for the 3d Ising gauge model in which, thanks to the efficiency of the existing  
Montecarlo algorithms for this model, 
large enough values of the interquark distance could be reached and unambiguous signatures of the logarithmic 
increase of the flux tube thickness (at zero temperature) could be observed~\cite{cgmv95}. 
 
In this paper we propose an alternative way to address the above question in the vicinity of the deconfinement 
transition using the Svetitsky-Yaffe conjecture which is a very powerful tool  
to study the finite T behaviour of a confining LGT in the vicinity of the deconfinement point, at least   
for those LGT's whose  deconfinement transition is of second order. The Svetitsky-Yaffe conjecture~\cite{sy82} 
states that the deconfinement transition of a (d+1) LGT lies  
in the same universality class of the magnetization transition 
of a d-dimensional spin model with symmetry group given by the center of the original gauge group. 
 
This gives us a non trivial opportunity to check the effective string predictions.  
If we choose a (2+1) dimensional LGT with a 
gauge group with center $Z_2$ (like the gauge Ising model or the SU(2) or SP(2) LGT's which all have continuous 
deconfinement transitions), the target spin model is the 2d Ising model in the high temperature symmetric phase 
for which several exact results are known. In particular we shall see that it is possible to study analytically 
the equivalent of the flux tube thickness. Remarkably enough the results that we find agree with the effective 
string ones thus strongly supporting the idea of a smooth transition from a log to a linear behaviour as the 
temperature increases.  
 
This paper is organized as follows. Sect.2 is devoted to the discussion of some background material on the 
effective string description of the flux tune thickness and on the Svetitsky Yaffe conjecture. In sect.3 we 
study the $\langle \sigma_1 \epsilon_2 \sigma_3 \rangle$ correlator in the 2d Ising model, discuss its asymptotic 
behaviour and extract from it a prediction for the flux tube thickness. Finally in sect.4 we compare it with the 
effective string prediction and discuss some further features of our analysis.

\section{Background} 
 
\subsection{Definition of the flux tube thickness} 
 
The lattice operator which is commonly used to evaluate the flux density in presence of a pair of Polyakov loops 
(or equivalently in presence of a Wilson loop) is the following correlator 
\eq 
<\phi(x_0,x_1,x_2,R)>=\frac{<P(0,0)P^+(0,R)U_p(x_0,x_1,x_2)>}{<P(0,0)P^+(0,R)>}-<U_p> 
\label{flux} 
\en 
where $P(x_1,x_2)$ denotes a Polyakov loop in the spacelike position $(x_1,x_2)$ (in the above equation we have 
chosen for simplicity to locate the Polyakov loops in the positions (0,0) and (0,R) and the $x_0$ coordinate runs 
in the compactified timelike direction ) while $U_p(x)$ denotes a 
plaquette located in $x\equiv(x_0,x_1,x_2)$. The different possible orientations of the plaquette measure the 
different components of the flux. In the following we shall neglect this dependence (see however the comment at 
the end of the next section). In order to avoid boundary 
effects we then concentrate on the midpoint of the Polyakov loop correlator choosing a generic value of $x_0$ and 
fixing $x_2=R/2$. With this choice the flux density will be only function of the interquark distance $R$ and of 
the transverse coordinate $x_1$. In the $x_1$ direction the flux density shows a gaussian like shape (see for 
instance fig.2 of ref.~\cite{cgmv95}). The width of this gaussian  
is the quantity which is usually denoted as ``flux tube thickness'': $w(R,N_t)$ . This quantity only depends on 
the interquark distance $R$ and  on the lattice size in the compactified timelike 
direction $N_t$, i.e. on the inverse temperature of the model  
(this dependence was implicit in  the above definition). 
By tuning $N_t$ we can thus study the flux tube thickness in the vicinity of the deconfinement transition 
 
\subsection{Dimensional reduction and the Svetitsky Yaffe conjecture.} 
Given a (2+1) LGT with gauge group $G$ and a two dimensional spin model with symmetry group the center of $G$, 
if 
 both the  
deconfinement transition of the  
(2+1) dimensional LGT and the magnetization transition of the 2d spin model are continuous 
then, according to the Svetitsky--Yaffe conjecture~\cite{sy82}, the two critical points must  
belong to the same universality class and we can use the spin model as an effective theory description for the  
(2+1) dimensional LGT in the neighbourhood of the deconfinement transition. This is the case if we choose for instance $SU(2)$ as gauge group for the (2+1)  
dimensional  
LGT and $Z_2$ (the center of $SU(2)$) as symmetry group of the 2d spin model (i.e. the Ising model). Another 
equivalent choice would be to study the (2+1) dimensional LGT with gauge group $Z_2$ (the Ising gauge 
model) whose 
center obviously is again $Z_2$. 
 
In this effective description the Polyakov loops of the LGT are mapped onto the spins of the Ising  
model, the confining phase of the LGT into the high temperature phase of the spin model while the combination  
$\sigma(N_t) N_t$ (where $\sigma(N_t)$ denotes the finite temperature value of the string tension while in the 
following $\sigma$ with no explicit dependence will denote by default the zero temperature string tension)  
is mapped into the  
mass scale of the  
spin model (i.e.the inverse of the correlation length) and sets the scale of the deviations from the critical  
behaviour. In order to describe in the dimensionally reduced model the expectation value of eq.(\ref{flux})  
one must extend this mapping also to 
the plaquette operator. This non-trivial problem was discussed a few years ago 
by Gliozzi and Provero in~\cite{gp97}, where they were able to show that   
in the vicinity of the critical (deconfinement) 
 point the plaquette operator of the (2+1) LGT is mapped into a mixture of the energy and identity operators 
of the 2d spin model. As a consequence the combination of gauge invariant operators which measures  
the density of 
chromoelectric flux in a meson is mapped into a three point function 
$ <\epsilon\sigma\bar\sigma>$ where in the Ising case (i.e. for LGT with a $Z_2$ center symmetry) 
$\bar\sigma=\sigma$. Different components of the flux (i.e. different orientations of the plaquette)  
correspond to different coefficients in the linear combination which relates the plaquette operator of the LGT 
with the energy and identity operators of the spin model. These coefficients will play no role in the 
discussion and we shall neglect them in the following.

\subsection{Agreement  between 2d spin model estimates and effective string predictions.} 
If we describe  
the correlator of two Polyakov loops in the high temperature regime of a confining 
lattice gauge theory in d=3 using the Nambu-Goto effective string we obtain exactly a collection of $K_0$ Bessel 
functions. This was observed for the first time by L\"uscher and Weisz~\cite{lw04} using a duality transformation 
and then derived in the covariant formalism in~\cite{bc05}. The expression which one obtains is: 
\eq 
  \left\langle P(0,0) P(0,R)^+\right\rangle 
  =\sum_{n=0}^{\infty}\left|v_n\right|^2 
  2R\left({\tilde{E}_n\over2\pi R}\right)^{{1\over2}(d-1)} 
  K_{{1\over2}(d-3)}(\tilde{E}_nR). 
\en 
(see eq. (3.2) of~\cite{lw04}). 
Notice that in $d=3$ the $R$ dependence in front of the Bessel functions cancels exactly.  
 
The argument of the $K_0$ functions is given by the product of the interquark distance $R$ and the  
closed string energy levels: 
  
 \eq 
  \tilde{E}_n=\sigma T 
  \left\{1+{8\pi\over\sigma N_t^2}\left[-\frac{1}{24}\left(d-2\right)+n\right] 
  \right\}^{1/2}. 
  \en 
  (see eq. (C5) of~\cite{lw04}).

It is easy to see that in the large $R$ limit only the lowest state $(n=0)$ survives and we end up with a single 
$K_0$ function: 
\eq 
\lim_{R\to\infty}  \left\langle P(0,0) P(0,R)^+\right\rangle 
  \sim  
  K_{0}(\tilde{E}_0R). 
\label{rtoinfty} 
\en 
 
where: 
\eq 
  \tilde{E}_0=\sigma N_t 
  \left\{1-{\frac{\pi}{3\sigma N_t^2}} 
  \right\}^{1/2}. 
  \en 
 
From this expression we read the Nambu-Goto prediction for the finite temperature dependence of the string 
tension: 
\eq 
  \sigma(N_t)\equiv \tilde{E}_0/N_t=\sigma  
  \left\{1-{\frac{\pi}{3\sigma N_t^2}} 
  \right\}^{1/2}. 
  \en 
As it is well known this expression cannot be exact since it predicts mean value critical indices for the 
deconfinement transition, however it turns out to be a very good approximation up to rather high temperatures (we 
shall further comment on this below). 
 
Looking at eq.(\ref{rtoinfty}) 
it is tempting to identify the Polyakov loops correlator in the large $R$ 
 limit with the spin-spin correlator in the 2d Ising model  
which is exactly given by a $K_0$ Bessel function with argument $m R$, $m$ being the mass of the Ising 
model. This is the origin of the relation mentioned above between the mass of the 
effective Ising model: 
 $m$ and  the product $\sigma(N_t)N_t$ of the 
LGT. Notice as a side remark that this correspondence is more general than this particular Ising case, since any 
2d spin model with a spectrum which starts with an isolated pole has a spin-spin correlator dominated at large 
distance by a $K_0$ Bessel function without prefactors. 
 
A few comments are in order at this point: 
\begin{itemize} 
\item 
The experience with Polyakov loop correlators~\cite{es1,es2,es3,es4}  
shows that the Nambu-Goto action is a good approximation (indeed a 
very good one) for very large $R$ (much larger than $N_t$ ) and values of $N_t$ such that:  
$N_t\geq \sqrt{4/\sigma}$. For higher temperatures (i.e. smaller values of $N_t$) the deviations due to the ``mean 
field '' nature of the Nambu-Goto approximation cannot be neglected.  
 
\item 
It will be useful in the following to define the combination  $t_f\equiv \frac{1}{N_t\sqrt{\sigma}}$.  
With this definition 
$t_f$ is dimensionless and has the meaning of a finite temperature.  
The Nambu-Goto approximation would suggest a deconfinement transition for 
$t_f=\sqrt{\pi/3}\sim 1.02$ while the bound $N_t\geq \sqrt{4/\sigma}$ mentioned above corresponds to $t_f<0.5$.  
It is interesting to observe that the Nambu-Goto prediction for the critical temperature is different, but non 
too far, from the known 
Montecarlo estimates for LGT's: in the gauge Ising model the 
deconfinement transition occurs at $t_f\sim 1.2$~\cite{ch96} while in the $SU(2)$ case we have $t_f\sim 
1.13$~\cite{t98}. 
 
\item 
Looking at $\tilde E_0$ we see that it is useful to define the dimensionless coefficient 
\eq 
\rho\equiv  \left\{1-{\frac{\pi t_f^2}{3}}\right\}^{1/2} 
\en 
from which we find 
\eq 
m=\sigma T\rho 
\en 
The region ($t_f\leq 0.5$) in which we can trust the Nambu-Goto approximation  
corresponds to the range $0.8\leq\rho\leq 1$

\end{itemize} 
 
\subsection{Effective string predictions for the flux tube thickness.}

The simultaneous dependence of the flux tube thickness on the two variable $R$ and $N_t$ 
 can be evaluated exactly only 
in the gaussian limit. Including higher order terms in the effective string action makes the problem too 
difficult (even if some recent result in the framework of the covariant quantization suggest that some 
simplification could occur if one chooses to study the whole Nambu-Goto action \cite{bc05,bcf06}). 
 
For the details of the calculations we refer the reader to the paper~\cite{cgmv95} (see also~\cite{p99}) 
For our purpose we are only interested in the two asymptotic limits: large $N_t$ and finite $R$  
(which is the zero 
temperature limit where we expect a log type behaviour) and the opposite one: large $R$ and small $N_t$ which is 
high temperature limit.  
 
One finds: 
\eq 
w^2\sim\frac{1}{2\pi\sigma}\log(\frac{R}{R_c})~~~~~~~~~~~(N_t>>R>>0)  
\en 
 
\eq 
w^2\sim\frac{1}{2\pi\sigma}(\frac{\pi R}{6 N_t}+\log(\frac{N_t}{2\pi}))~~~~~~~~~~~(R>>N_t)  
\label{espred} 
\en 
As it is easy to see in the second limit the logarithmic dependence is on $N_t$  
(the inverse of the temperature) and 
not on $R$ which appears instead in the linear correction.

\section{The 3-point correlators in the 2d Ising model} 
As we have seen the study of the width of long color flux tube in  
the (2+1)d LGT's with gauge group $Z_2,SU(2)$ or $Sp(2)$ can be translated  
in the study of the ratio of correlators 
\bea 
\frac{\langle \sigma(x_1) \epsilon(x_2) \sigma(x_3)\rangle} 
{\langle \sigma(x_1) \sigma(x_3)\rangle} 
\eea 
 
 in the high temperature phase 
 of the 2d Ising model in zero magnetic field. 

Since we are interested in the large distance behaviour of such a quantity, we will use the so-called Form Factors technique (see \cite{Yurov} for its application in the context of the 2d Ising model without magnetic field). 
Form factors are defined as suitable matrix elements of an operator $\phi$ between the vacuum and an arbitrary $n-$particle asymptotic state\footnote{We use the parametrization of energy and momentum in terms of the rapidity $\theta$: $E= m \cosh \theta$, $P= m \sinh \theta$.} 
\bea 
F^{\phi}_{a1 \dots a_n} (\theta_1, \dots, \theta_n) =  
\langle 0 | \phi(0) | A_{a_1}(\theta_1) \dots  A_{a_n}(\theta_n)\rangle. 
\eea 
When the theory is integrable, and the $S-$matrix is exactly known they can be computed exactly as solutions of certain functional equations \cite{Karowski,Smirnov}. In the present case the theory is free and the $S-$matrix is simply $S=-1$. 
 
Such a technique turned out to be an effective tool in order to give approximate expressions for the large distance behaviour of the two-point correlators in Integrable QFTs, and recently it has been employed to analyze the three-spin correlator in the 3-states Potts model \cite{cdgjm05}.
In brief, it is possible to rewrite a generic correlator as a spectral expansion whose building blocks are just the form factors. For example, for the connected two-point correlator we have in general 
\bea 
\langle \phi(x) \phi(0) \rangle_c = \sum_{n=1}^{\infty} \frac{1}{n!} \int_{-\infty}^{\infty} \frac{d \theta_1 \dots d \theta_n}{(2 \pi)^n} \, |F_n^\phi (\theta_1, \dots, \theta_n)|^2 \; e^{-m |x| E_n} 
\eea 
where $E_n$ is the energy of the $n-$particles state as function of the rapidity.

\begin{figure} 
\centering 
\includegraphics[height=7cm]{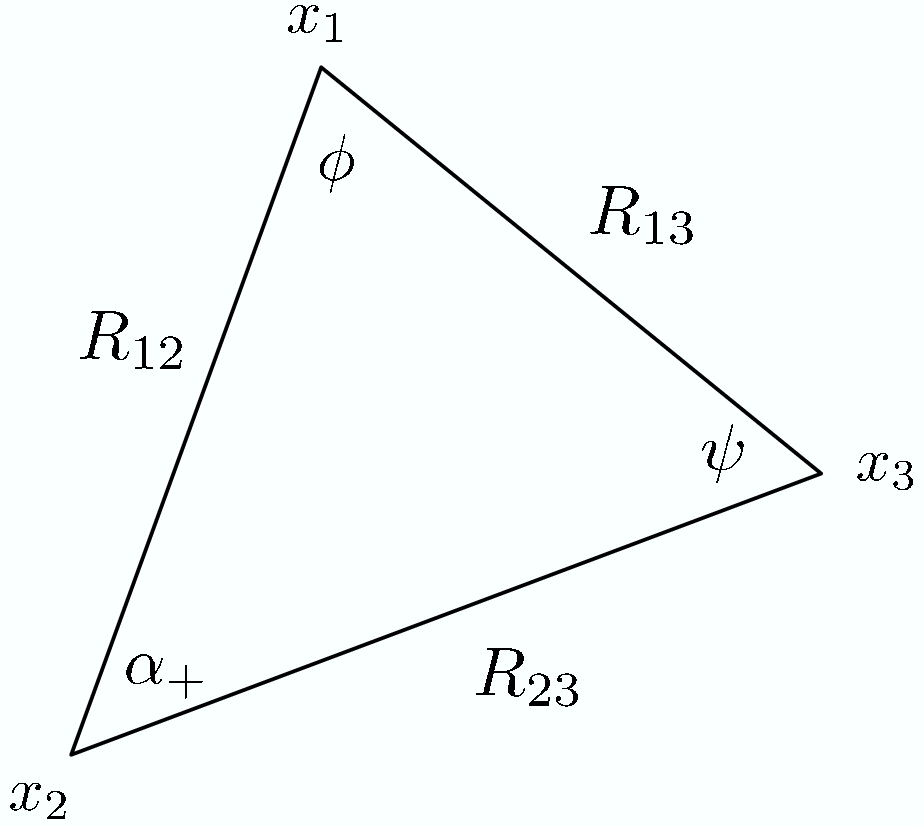} 
\vskip0.5cm 
\caption{Generic configuration for the three-point function $\langle \sigma(x_1) \epsilon(x_2) \sigma(x_3)\rangle$.} 
\label{fig3p} 
\end{figure}

Similar expressions can be written for $n-$points connected correlators, in particular for the three-point correlation function we can proceed in 
close analogy with the analysis of \cite{cdgjm05}. Hence we can write (see fig.\ref{fig3p}): 
\bea 
&& \langle \sigma(x_1) \epsilon(x_2) \sigma(x_3) \rangle  =  
\int_{-\infty}^{\infty} \frac{d \theta_1 \, d \theta_2}{ (2 \pi)^2} \, (F_1^\sigma)^2 \, F_2^\epsilon  
(\theta_{12} + i \alpha_+) \, e^{- m (R_{12} \cosh \theta_1 \, + \, R_{23} \cosh \theta_2) } + \nonumber \\ 
&& + \frac12 \int_{-\infty}^{\infty} \frac{d \theta_1 \, d \theta_2 \, d \theta_3}{ (2 \pi)^3} \, 
F_1^\sigma \, F_2^\epsilon(\theta_{23}) \, F_3^\sigma(\theta_{32}, \theta_{21}+ i \psi , \theta_{31}+ i \psi) \, 
 e^{- m (R_{13} \cosh \theta_1 \, + \, R_{23}( \cosh \theta_2 +\cosh \theta_3 )) } + \nonumber \\  
&& + \frac12 \int_{-\infty}^{\infty} \frac{d \theta_1 \, d \theta_2 \, d \theta_3}{ (2 \pi)^3} \, 
F_1^\sigma \, F_2^\epsilon(\theta_{12}) \, F_3^\sigma(\theta_{21}, \theta_{31}+ i \phi , \theta_{32}+ i \phi) \, 
 e^{- m (R_{13} \cosh \theta_3 \, + \, R_{12}( \cosh \theta_1 +\cosh \theta_2 )) } + \dots  
\label{tricorr} 
\eea 
The main ingredients which enter the previous formula are the first few form factors of the operators $\sigma$ and $\epsilon$. Let us briefly review their properties (see \cite{Yurov}). 

\underline{Spin $\sigma$ and disorder $\mu$ operators} 
 
Since we are dealing with a theory of free Majorana fermions, the $S-$matrix is simply $S=-1$. As a consequence of the fact that we are in the high-T phase of the theory, symmetry implies that the form factors of $\sigma$ and $\mu$ are non-zero upon odd and even particle states respectively.   
 
The VEV of $\mu$ is known since a long time and happens to be 
\bea 
\langle \mu \rangle = F_0^\mu = B |\tau|^{1/8} = \frac{B}{(2 \pi)^{1/8}} m^{1/8} = C \, m^{1/8}, \ \ \ B=1.70852190 \dots                                    
\eea  
where we used the exact relation between the coupling constant $\tau$ (the reduced temperature) and the mass of the fermion $m= 2 \pi \tau$.  
The cluster condition \cite{Delfino:1996nf} fixes the relative normalization between the FF of $\sigma$ and $\mu$, and implies $F_1^\sigma = \langle \mu \rangle$.  
The explicit expression of $F_3^\sigma$ is given by 
\bea 
F_3^\sigma(\theta_1, \theta_2, \theta_3) = i \, F_1^\sigma \, \prod_{i<j=1}^3 \, \tanh \frac{\theta_{ij}}{2} 
\eea 
and the normalization constant can be fixed by means of the residue equation on the annihilation poles 
\bea 
- i \,{ \textrm{Res} }_{\theta_{12} = i \pi}   F_3^\sigma(\theta_1, \theta_2, \theta_3) \ = \  (1-S) \, F_1^\sigma  = 2 \,  F_1^\sigma. 
\eea 
 
\underline{Stress-energy tensor $\Theta$ and energy density $\epsilon$} 
 
The FF of the perturbing operator $\epsilon$ can be extracted from those of the stress-energy tensor $\Theta$. 
Their relationship is given by (we recall that $\Delta_\epsilon = 1/2$) 
\bea 
\Theta \ = \  4 \pi (1- \Delta_\epsilon) \, \tau \, \epsilon = 2 \pi \tau \, \epsilon = m \, \epsilon.  
\eea 
The fact that we have $S=-1$ implies that the only non-zero form factor of the trace $\Theta$ is the two-particle one $F_2^\Theta$  
\bea 
F_2^\Theta (\theta) \ = \ -2 i \pi m^2 \, \sinh \frac{\theta}{2}   
\eea 
which have been normalized by means of the condition $F_2^\Theta (i \pi ) = 2 \pi m^2$.  
The final step is to write $F_2^\epsilon$ using the relation $\Theta = m \epsilon$ 
\bea 
F_2^{\epsilon} = \frac{1}{m} F_2^{\Theta} \ = \ -2 i \pi m \, \sinh \frac{\theta}{2}. 
\eea 
All the other FFs of $\epsilon (x)$ and $\Theta(x)$ are zero. 
\subsection{The width of the flux tube} 
We are now in the position to give an analytic estimate for the width of the flux tube in the limit of large distances. 
When the sides of the triangle are large, the leading behaviour of (\ref{tricorr}) is given by the first term in the rhs.  
It is interesting to notice that such a term can be written in an explicit way for an arbitrary triangle. The reason lies in the simple form of the 2p form factor of the energy operator which allows to write it as follows 
\bea 
&& \int_{-\infty}^{\infty} \frac{d \theta_1 \, d \theta_2}{ (2 \pi)^2} \, (F_1^\sigma)^2 \, F_2^\epsilon  
(\theta_{12} + i \alpha_+) \, e^{- m (R_{12} \cosh \theta_1 \, + \, R_{23} \cosh \theta_2) } = \nonumber \\ 
&&= m  \, (F_1^\sigma)^2 \sin \frac{\alpha_+}{2} \int_{-\infty}^{\infty} \frac{d \theta_1 \, d \theta_2}{ 2 \pi} \,  
\cosh \frac{\theta_1}{2} \, \cosh \frac{\theta_2}{2}  
\, e^{- m (R_{12} \cosh \theta_1 \, + \, R_{23} \cosh \theta_2) } 
= \nonumber \\ 
&&= \, (F_1^\sigma)^2 \sin \frac{\alpha_+}{2} \frac{e^{- m (R_{12}+ R_{23})} }{\sqrt{R_{12}R_{23}}} 
\eea 
and hence 
\bea 
\langle \sigma(x_1) \epsilon(x_2) \sigma(x_3) \rangle  =  (F_1^\sigma)^2 \, 
\frac{\sqrt{R_{13}^2-(R_{12}-R_{23})^2}}{2\,R_{12}R_{23}} \, 
e^{- m (R_{12}+ R_{23})} + \dots 
\eea 
Since we are interested in analyzing the width of the flux tube at the mid-point between $\sigma (x_1)$ and 
 $\sigma (x_3)$ for large separations of them, we will put $R_{12} = R_{23} =L$ and $R_{13} = 2 r $ in 
  (\ref{tricorr}), and we define the transverse distance as $y =L \cos \alpha_+/2$  
  (we also have $r=L \sin \alpha_+ /2$). Then, we define the three-point function in such a geometric configuration as $S(r,y)$  
\bea 
S(r,y) \ = \  (F_1^\sigma)^2 \, \frac{r}{y^2+r^2} \, e^{-2 m \sqrt{y^2+r^2}} . 
\label{formfig}
\eea 
 
In order to study the flux tube shape we must study the following ratio 
\bea 
P(r,y) \ = \ \frac{S(r,y)}{\langle \sigma(2r) \sigma(0)\rangle} 
\eea 
where $\langle \sigma(2r) \sigma(0)\rangle$ is the two-point correlator between $\sigma(x_1)$ and $\sigma(x_3)$. In the large distance limit its leading behaviour is  
\bea 
\langle \sigma(2r) \sigma(0)\rangle \ \simeq \ \int_{-\infty}^{\infty} \frac{d \theta}{ 2 \pi } (F_1^{\sigma})^2 \, e^{-2mr \, \cosh \theta} 
= \frac{(F_1^\sigma)^2}{\pi} K_0(2 mr)  
\eea 
and hence the ratio can be cast in the form 
\bea 
P(d,y) \ = \ \frac{S(r,y)}{\langle \sigma(2r) \sigma(0)\rangle} = \frac{ \pi \, r}{y^2+r^2} \, \frac{e^{-2 m \sqrt{y^2+r^2}}}{K_0(2mr)}.  
\label{nongaussian} 
\eea 
It is easy to see that for very large separations $(m r \to \infty)$ the shape becomes of the gaussian type in the transverse variable $y$ 
\bea 
P(r,y) \ \simeq \  \frac{ \pi \, r}{y^2+r^2} \, \frac{e^{-2 m r }}{K_0(2mr)} \, e^{-\frac{m}{r} \, y^2}.  
\eea 
Finally we can study the variance $w^2(r)$ of $P(r,y)$ wrt $y$, which exactly corresponds to the   
width of the flux tubes which we were looking for.  
Let us define: 
\bea 
w^2(r) \ =  \int_{-\infty}^{\infty} dy \, y^2 \, P(r,y)  
\eea 
as a consequence, setting $x=y/r$ we obtain 
\bea 
w^2(r) \ = \ \frac{ \pi r^2}{K_0(2mr)} \, \int_{-\infty}^{\infty} dx \, \frac{x^2}{1+x^2} e^{-2mr \sqrt{1+x^2}}. 
\eea 
Such an integration cannot be performed exactly, but we can still give its asymptotic estimate in the limit $mr \to \infty$.  
With standard techniques we obtain 
\bea 
\int_{-\infty}^{\infty} dx \, \frac{x^2}{1+x^2} e^{-2mr \sqrt{1+x^2}} \ \simeq \ 
\sqrt{\pi} \, e^{-2mr} 
\left(\frac12 (mr)^{-3/2} -\frac{9}{32}  (mr)^{-5/2} + O( (mr)^{-7/2} )\right). 
\eea 
and combining it with 
\bea 
\frac{1 }{K_0(2 mr)} \  \simeq  \ \frac{e^{2mr}}{\sqrt{\pi}}  
\left(2 (mr)^{1/2} +\frac{1}{8}  (mr)^{-1/2} + O( (mr)^{-3/2} )\right) 
\eea 
we finally obtain 
\bea 
w^2(mr) \ \simeq \  \pi\, \frac{r}{m} \,\left(1 - \frac{1}{2mr} + O\left(\frac{1}{(mr)^2}\right) \right) 
\eea 
which states that the width of the flux tube, in the limit of very large separations between the spins, behaves linearly with the separation $R = 2 r$  
\bea 
w^2(m R ) \ \simeq \  \frac{\pi}{2} \, \frac{R}{m} -\frac{\pi}{2m^2} + \dots. 
\label{result} 
\eea 
 
Finally, it is not difficult to calculate the asymptotic expansion of the higher order momenta  
\bea 
w^{(2n)}(r)=\frac{\pi \, r^{2n}}{K_0 (2mr)}\int_{-\infty}^{\infty} dx \, \frac{x^{2n}}{1+x^2} e^{-2mr \sqrt{1+x^2}}. 
\eea 
It turn out to be  
\bea 
w^{(2n)}(r)=\frac{\pi \, e^{-2 mr}}{\sqrt{mr} \, K_0 (2mr)} \left( \frac{r}{m} \right)^n \; 
\lim_{N \to \infty} \, \sum_{k=0}^N \sum_{j=0}^{N-k}  
\frac{(-1)^j}{4^k \, k!} \Gamma(2k+j+n+1/2) \, (mr)^{-k-j}, 
\label{hm1} 
\eea 
where the first few orders are as follows 
\bea 
w^{(2n)} (r) =  \sqrt{\pi} \, \Gamma(n+1/2) \left(\frac{r}{m}\right)^n \left(2+\frac{1}{mr} 
( n^2/2 -n-1/2) + O\left(\frac{1}{(mr)^2}\right) \right). 
\label{hm2} 
\eea 
 
\section{Results} 
 
\subsection{Comparison with the effective string predictions} 
 
Eq.(\ref{result}) is the main result of this paper. If we compare it with the effective string prediction of 
eq.(\ref{espred}) we see a remarkable agreement between the two results. 
Both show a linear increase with the interquark distance of the flux tube thickness. The linear term appears in 
both equations with the same  
 $R/N_t$ dependence and the right dimensions given by 
$\frac1\sigma$. The factor in front of this correction is 
not the same in the effective string and in the spin model cases, but this is not strange since there is a 
finite renormalization in the mapping between the plaquette operator (in the LGT) and the energy operator in the 
spin model (and similarly between the Polyakov loop and the spin operator). Moreover one must recall that the 
effective string result is obtained in  the framework of the purely gaussian approximation

\subsection{Isolines of chromoelectric flux} 
 
\begin{figure} 
\centering 
\includegraphics[width=16cm]{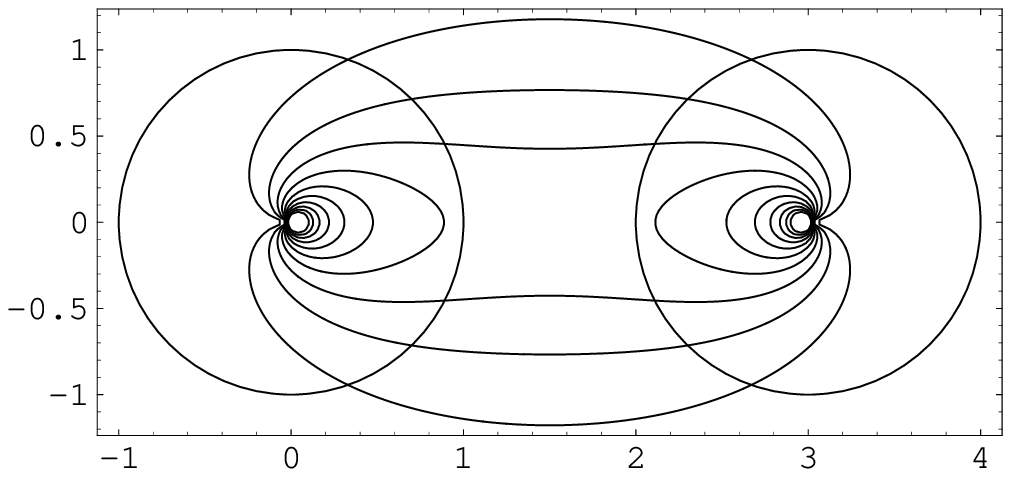} 
\vskip0.5cm 
\caption{Isolines of chromoelectric flux for $m=1$. The circles are centered in correspondence of the static quarks and their radius is $1/m$. Their extention gives a measure of the region where we expect large corrections to the leading behaviour we used to produce both the picture and the estimate of the fluxtube width (we recall that we expect that the leading behaviour is valid for $m \, R_{ij} >> 1$ where $R_{ij}$ is a generic side of the triangle in figure 1).  
\label{fig05} 
} 
\end{figure}

It is interesting to plot the isolines  of the chromoelectric flux as a function of the interquark distance in 
the approximation of the two dimensional Ising model.  
These can be easily obtained from eq.(\ref{formfig}). Since in the reduced model we have only one scale $m~l$  
which combines 
both the finite temperature and the interquark distance of the original LGT these plots can be interpreted in two 
ways. 
\begin{itemize} 
\item 
We may look at them as the result of keeping the temperature of the LGT fixed (i.e. $m$ fixed) 
then as $ml$ increases we are effectively describing an  increase of the interquark distance $R$. It is nice to 
see that indeed as $ml$ increases (from fig.\ref{fig05} to fig.\ref{fig10}) 
the shape of the flux tube becomes more and more narrow, as one should expect 
for a confining theory. 
\item 
Alternatively we may think to keep the interquark distance fixed and look to what happens as we increase the 
temperature and approach the deconfinement transition (i.e. as $m$ decreases). Again it is nice to see that the 
flux tube smoothly moves from the narrow shape of a confining theory (fig.\ref{fig10}) to a shape in which the flux 
lines become more and more delocalized in the two dimensional surface (fig.\ref{fig05}). 
\end{itemize}

\subsection{The shape of the flux tube.} 
Finally it is interesting to notice that from the explicit expression of the  
 $\langle \sigma_1 \epsilon_2 \sigma_3 \rangle$  correlator one can see  
that the shape of the flux tube is not exactly 
a gaussian (see eq.(\ref{nongaussian})).  
This is indeed also what is visible in the montecarlo simulations (see fig.2 of ~\cite{cgmv95} and the 
whole discussion at the beginning of \cite{mm04}). Our analysis could suggest a tentative analytic form for these 
deviations which should be valid in the neighborhood of the deconfinement transition and hence extend to this 
regime the functional form valid at low temperature proposed in~\cite{mm04}. This prediction is encoded in the  
higher order momenta reported in eq.s~(\ref{hm1},\ref{hm2}). 
It would be nice 
to check this prediction by performing also in the high temperature regime a set of 
 montecarlo simulations similar to those discussed in~\cite{mm04}.

\vskip1.0cm {\bf Acknowledgements.} The work of P.G. is supported by the European Commission TMR programme  
HPRN-CT-2002-00325 (EUCLID).  
 The authors would like to thank 
 F. Gliozzi, M. Bill\'o and L. Ferro for 
 useful discussions. 

\vskip1.0cm {\bf Note Added.} An interesting application of form factors to the (2+1)-d $SU(2)$ gauge theory \cite{Orland:2006wq} has appeared at the same time as the present paper. In such a work, the (2+1)-d $SU(2)$ gauge theory is generalized to an anisotropic form with two gauge couplings, and the form factors of the currents of the $SU(2)$ principal chiral model in (1+1)-d have been used to compute the string tension in the anisotropic regime. In this context, the mechanism of dimensional reduction is not related to the Svetitsky-Yaffe conjecture.

\newpage 
 
\begin{figure} 
\centering 
\includegraphics[width=16cm]{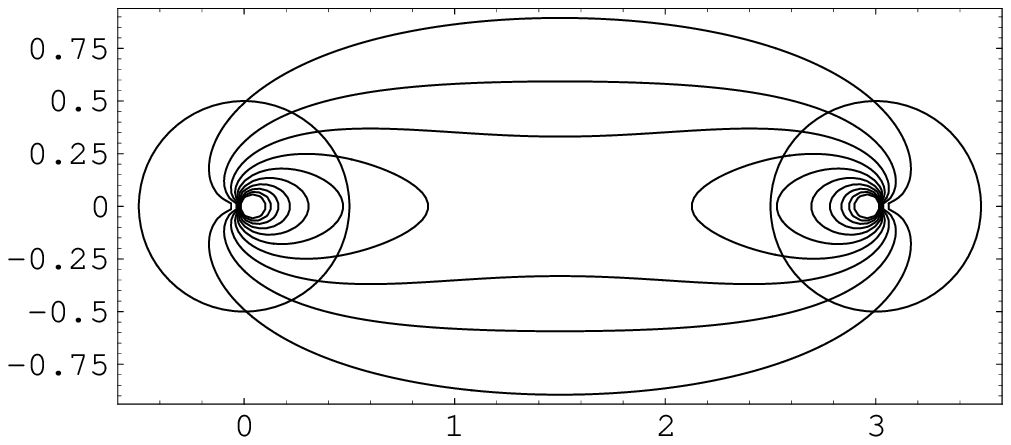} 
\vskip0.5cm 
\caption{Isolines of chromoelectric flux for $m=2$  
\label{fig1} 
} 
\end{figure}

\begin{figure} 
\centering 
\includegraphics[width=16cm]{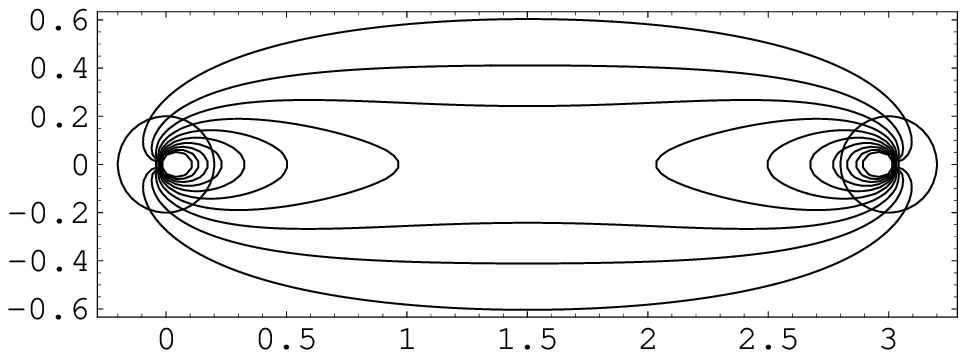} 
\vskip0.5cm 
\caption{Isolines of chromoelectric flux for $m=5$  
\label{fig5} 
} 
\end{figure}

\begin{figure} 
\centering 
\includegraphics[width=16cm]{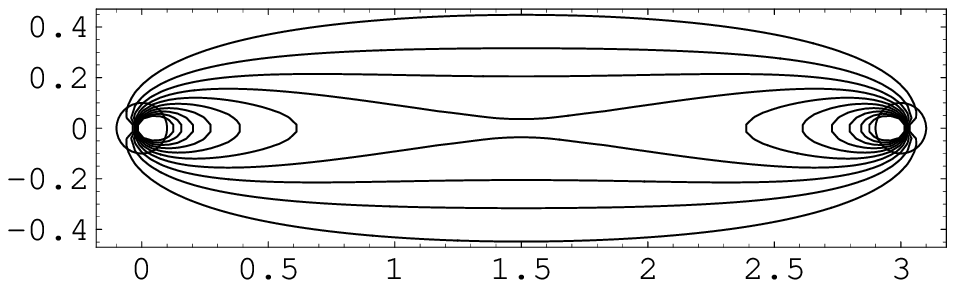} 
\vskip0.5cm 
\caption{Isolines of chromoelectric flux for $m=10$  
\label{fig10} 
} 
\end{figure}

\end{document}